\documentclass{aastex}
\usepackage{emulateapj5,natbib,multicol}
\include{epsf}
\renewcommand{\baselinestretch}{1.1}

\begin{document}
%
%

%
%
\pagestyle{plain}
\def\SZ{Sunyaev-Zel'dovich~}
\def\ie{{\em i.e.}}
\def\eg{{\em e.g.}}
\def\hou{km s$^{-1}$ Mpc $^{-1}$}
\def\omega0{$\Omega_\circ$}
\def\omegam{$\Omega_M$}
\def\omegal{$\Omega_\Lambda$}
\def\lambdao{$\Lambda_\circ$}
\def\Ho{$H_{\circ}$}
\def\rc{$\theta_c$}
\def\be{$\beta$}
\def\rms{$r.m.s.$}
\def\uv{$u$-$v$}
\def\r500{$r_{500}$}
\def\dt{$\Delta T(0)$}
\def\dchi{$\Delta\chi^2$}
\def\qo{$q_\circ$}
\def\da{$D_A$}
%
\newcommand{\vol}[1]{{\bf{ #1}}}
\newcommand{\nature}{\em Nature\rm}
\newcommand{\aanda}{{\em A. \& A.{\rm}}}
%
%
\newcommand{\ein}{{\em Einstein{\rm}}}
\newcommand{\exo}{{\em EXOSAT{\rm}}}
\newcommand{\ginga}{{\em Ginga{\rm}}}
\newcommand{\spek}{SPEKTROSAT}
\newcommand{\rosat}{{\em ROSAT}}
\newcommand{\asca}{{\em ASCA}}
\newcommand{\free}{Freestanding Instrument}
\newcommand{\hxt}{hard x-ray telescope}
\newcommand{\slew}{Slew Survey}
%
%
\newcommand{\logn}{{\em logN-logS{\rm}}}
\newcommand{\xrb}{x-ray background}
\newcommand{\dxrb}{diffuse x-ray background}
\newcommand{\aox}{\ifmmode{\alpha_{\tiny OX}} \else $\alpha_{\scriptsize OX}$\fi} 
\newcommand{\alpe}{\ifmmode{\alpha} \else $\alpha$\fi}
\newcommand{\atoms}{\ifmmode{\rm ~atoms~cm^{-2}} \else ~atoms cm$^{-2}$\fi}
\newcommand{\nh}{\ifmmode{\rm N_{H}} \else N$_{H}$\fi}
%
%
\newcommand{\kao}{Kuiper Airborne Observatory}
\newcommand{\iue}{\em International Ultraviolet Explorer}
%
%
\newcommand{\qed}{quasar energy distribution}
\newcommand{\qeds}{quasar energy distributions}
\newcommand{\ir}{infrared}
\newcommand{\optuv}{optical/ultraviolet}
\newcommand{\nufnu}{\ifmmode \nu f_{\nu} \else$\nu f_{\nu}$\fi}
\newcommand{\fnu}{\ifmmode f_{\nu} \else$f_{\nu}$\fi}
\newcommand{\sn}{signal-to-noise}
\newcommand{\ede}{E/$\Delta$E}
\newcommand{\agn}{active galactic nuclei}
\newcommand{\snr}{supernova remnants}
\newcommand{\mdot}{$\dot{M}$}
\newcommand{\msun}{$M_{\odot}$}
\newcommand{\rr}{$F_{PL} \over F_{disk}$}
\newcommand{\costheta}{$\cos \theta$}
\newcommand{\astar}{$a_{*}$}
\newcommand{\bapp}{\ifmmode \beta_{app}\else$\beta_{app}$\ \fi}
%
%
\newcommand{\erg}{\ifmmode erg~cm^{-2}~s^{-1}\else erg cm$^{-2}$ s$^{-1}$\fi}
\newcommand{\degs}{\ifmmode ^{\circ}\else$^{\circ}$\fi}
\newcommand{\pmn}{\ifmmode \pm\else$\pm$\fi}
\newcommand{\per}[1]{\ifmmode \^{#1}\else$^{#1}$\fi}
\def\farcm{\hbox{$.\mkern-4mu^\prime$}}
\def\farcs{\hbox{$.\!\!^{\prime\prime}$}}
\def\fdg{\hbox{$.\!\!^\circ$}}
%
%
\newcommand{\cf}{{\em c.f.}}
\newcommand{\vs}{{\em vs.}}
\newcommand{\via}{{\em via}}
\newcommand{\esp}{{\em esp.}}
\newcommand{\etc}{{\em etc.}}
\newcommand{\etal}{{\em et~al.}}
\newcommand{\twid}{\ifmmode\sim \else${\sim}$\fi}
\newcommand{\lapprox}{_<\atop^\sim}  
\newcommand{\gapprox}{_>\atop^\sim}  
\newcommand{\my}{mas~yr$^{-1}$ }

\input{psfig}
\title{Galaxy Cluster Gas Mass Fractions from Sunyaev-Zel'dovich
Effect Measurements: Constraints on \omegam}

\author{Laura Grego\altaffilmark{1}, John
E. Carlstrom\altaffilmark{2},  Erik
D. Reese\altaffilmark{2}, Gilbert P. Holder\altaffilmark{2}, William
L. Holzapfel\altaffilmark{5}, Marshall
K. Joy\altaffilmark{3}, Joseph J. Mohr\altaffilmark{2}, \& Sandeep Patel\altaffilmark{4}}\hfill\break
\altaffiltext{1}{Harvard-Smithsonian Center for Astrophysics, 60
Garden St., Cambridge, MA 02139}
\altaffiltext{2}{Department of Astronomy \& Astrophysics, 5640 S. Ellis Ave., University of Chicago, Chicago, IL 60637}
\altaffiltext{3}{Dept. of Space Science, SD50, NASA Marshall Space Flight Center, Huntsville, AL 35812}
\altaffiltext{4}{Department of Physics, University of Alabama, Huntsville, AL 35899}
\altaffiltext{5}{Department of Physics, University of California, Berkeley, CA 94720}
\authoraddr{Harvard-Smithsonian Center for Astrophysics, MS 83, 60
Garden St., Cambridge, MA 02139}

\begin{abstract}
Using sensitive centimeter-wave receivers mounted on the Owens Valley 
Radio Observatory and Berkeley-Illinois-Maryland-Association 
millimeter arrays, we have obtained interferometric measurements of 
the Sunyaev-Zel'dovich (SZ) effect toward massive galaxy clusters. 
We use the SZ data to determine the pressure distribution of the 
cluster gas and, in combination with published X-ray temperatures, to 
infer the gas mass and total gravitational mass of 18 clusters.  The 
gas mass fraction, $f_g$, is calculated for each cluster, and is extrapolated 
to the fiducial radius $r_{500}$ using the results of numerical
simulations.  The mean $f_g$ within $r_{500}$ is 
$0.081^{+0.009}_{-0.011}h^{-1}_{100}$ (statistical uncertainty at 
68\% confidence level, assuming \omegam=0.3, \omegal=0.7).  We discuss possible sources of systematic 
errors in the mean $f_g$ measurement.

We derive an upper limit for \omegam\ from this sample under 
the assumption that the mass composition of clusters
within $r_{500}$ reflects the universal mass composition: $\Omega_M h \le
\Omega_B/f_g$.  The gas mass fractions depend on cosmology through the
angular diameter distance and the $r_{500}$ correction factors.
  For a flat universe (\omegal $\equiv$ 1 - \omegam) and $h=0.7$, we find the
measured gas mass fractions are consistent with \omegam\ less than
0.40, at 68\% confidence.  Including estimates of the baryons contained in 
galaxies and the baryons which failed to become bound during 
the cluster formation process, we find \omegam\ $\sim$0.25. 

\end{abstract}

\section{Introduction}
\label{chap:intro}
Clusters of galaxies, by virtue of being the largest known virialized objects, 
are important probes of large scale structure and can be used to test cosmological models.  
Rich clusters are extremely massive, $\sim$$\ 10^{15} M_{\odot}$, as indicated by the presence of strongly
 gravitationally lensed background galaxies, the large velocity dispersion ($>$ 1000 km s$^{-1}$) in the member galaxies,
 and the high measured temperature ($> 5$ keV) of the ionized
intracluster gas.  
The mass composition on cluster mass scales is expected to reflect the
universal mass composition \citep{white1993,evrard1997}
.  
Under the fair sample assumption, then, the cluster gas mass fraction, which is
a lower limit to the cluster's baryon fraction, $f_B$, should reflect the
universal baryon fraction:
\begin{equation} f_g \le f_B \equiv {\Omega_B \over \Omega_M},\end{equation}
where $\Omega_B$ is the ratio of baryon mass density in the universe
to the critical mass density.  The cluster gas mass fraction
measurement can be used within the Big Bang Nucleosynthesis (BBN)
paradigm to constrain $\Omega_M$,
\begin{equation}\Omega_M \le \Omega_B/f_g.\label{eq:omegam}\end{equation}
The value of $\Omega_B$ is constrained by BBN calculations and the
measurements of the abundances of the light elements \citep{wagonerfowlerhoyle,copischrammturner} as well as measurements of the spatial anisotropies of the
cosmic microwave background (CMB) \citep{whitescottsilk,hu97b}.  

The luminous baryons in clusters are mainly in the gaseous intracluster medium (ICM).  The gas mass is about an order of magnitude 
larger than the mass in optically observed cluster galaxies, \eg, \cite{white1993,formanjones1982}.  Hence, the gas mass is not only a lower 
limit to the cluster's baryonic mass, it is a reasonable estimate of
it.  Although observations suggest that galaxy
groups and low mass clusters may have lost gas due to preheating or
post-collapse energy input \citep{djf95,mohr1999a,ponman96}, the gas mass fraction in 
massive clusters ($T_e > 5$ keV) appears to be constant.  The gas mass
fraction in massive clusters then provides a lower limit to the cluster baryon fraction, $f_g \le f_B$.

 The ICM is hot, with electron temperatures, $T_e$, from $\sim$5 to 15 keV; rarefied, with peak electron 
number densities of $n_e \simeq 10^{-3}$ ${\rm cm}^{-3}$; and cools
slowly ($t_{cool}> t_{Hubble}$), mainly via thermal Bremsstrahlung
in the X-ray band. The ICM also produces a spectral distortion of the
CMB known as the Sunyaev-Zel'dovich effect.  The ICM mass fraction may
be calculated from either of these observables.  In addition to providing
measurements of this important parameter with independent techniques,
the two methods are fundamentally different in that the SZ effect is
directly proportional to the integrated density of the gas while the X-ray
 brightness is
proportional to the integrated square of the density.

The X-ray surface brightness is proportional
to the emission measure, $S_x \propto \int n_e^2 \Lambda(T_e) dl$, 
where the integration is along the
line of sight.  Under simplifying assumptions, the gas mass can be
calculated from an X-ray image deprojection.
Since the sound crossing time of the cluster gas is typically much
less than the dynamical time, one may reasonably assume that, in the
absence of a recent merger, the cluster gas is
relaxed in the cluster's potential.  Hydrodynamic simulations 
support this notion, \eg, \cite{EMN96}.  
Under the assumption that the gas is in hydrostatic equilibrium (HSE),
supported only by thermal pressure, the total binding mass follows
from the gas density and temperature distribution, the latter of which
may be determined with X-ray spectra.
Gas mass fractions have been measured with this technique 
out to cluster radii of 1 Mpc or more \citep{whitefabian95,djf95,neumann1997,squires1997,mohr1999a}.  In an X-ray
flux-limited sample of 45 clusters, \cite{mohr1999a} measure the mean cluster gas mass fraction
within approximately the virial radius to be (0.0749 $\pm 0.0005) h^{-3/2}$.  Here, and throughout the
paper, we use $H_\circ=100h$ km s$^{-1}$ Mpc$^{-1}$.

In this work, we calculate cluster gas mass fractions using spatially
resolved measurements of the SZ effect to determine the gas density
profile.  The SZ effect in clusters is a spectral distortion of the CMB radiation due to 
inverse-Compton scattering of the relatively cool CMB photons off hot
ICM electrons \citep{sunyaev1972}. 
At frequencies less than $\sim$~ 218 GHz, the intensity of the CMB radiation
is diminished as compared to the unscattered CMB, and the SZ effect
is manifested as a brightness temperature decrement towards the cluster.  This decrement, $\Delta T_{SZ} / T_{CMB}$,  
 has a magnitude proportional to the Compton $y$-parameter, \ie, the total
number of scatterers, weighted by their associated temperature, 
\begin{equation} y = {{k \sigma_T}\over {m_e c^2}} \int {n_e(l) T_e(l) dl}, \label{eq:compton}\end{equation}
where $k$ is Boltzmann's constant,  $\sigma_T$ is the Thomson
scattering cross section, $m_e$ is the electron mass. We extract the cluster's gas mass fraction 
from a deprojection of the SZ effect data in a method analogous to the
described X-ray HSE analysis.

In addition to providing an additional measurement of $f_g$, we 
note several points of difference between the X-ray and SZ analyses. 
Should significant large-scale or spatially-varying clumping of the 
ICM be present, the SZ 
image deprojection may look quite different from the X-ray
deprojection.  Clumping at scales below the resolution of the X-ray
and SZ images could also result in a difference of
$<n_e^2>^{1/2}/<n_e>$ in the inferred gas mass.   Also, the emission from the cores of 
relaxed clusters may be dominated 
by cooling flows, which complicate the interpretation of the X-ray data and
may bias the result strongly if not taken into account \citep{allen1998,mohr1999a}.  The direct relationship between the SZ effect and the
gas density also permits a surface gas mass fraction to be measured without
image deprojection by comparing the projected or ``surface''
gas mass from the SZ observation to a measurement 
of the surface total cluster mass, for instance from gravitational
lensing models (cf\ \cite{grego2000a}).
Because lensing observations are not available for all the clusters in
our sample, and
because we are interested in the gas mass fraction within clearly defined
cluster radii, for this work we calculate $f_g$ with the deprojection/HSE method only.

In this paper, we present cluster gas mass fractions 
based on SZ measurements made in the years 1994-1998, and a discussion of
the implications of these measurements for cluster physics and for
cosmology.  In previous papers \citep{carlstrom1996,grego2000a}, we have described
the instrument constructed expressly to make such measurements, and
the reduction and calibration methods for the SZ measurements.  We
give a brief review of this and discuss the cluster selection and observations in Section~\ref{sec:instrument}.   In
Section~\ref{sec:fgmethods}, we describe the procedure for fitting the
SZ data to models for the cluster gas and extracting cluster gas
masses and gas mass fractions, including a discussion of the possible
systematic uncertainties.  In Section~\ref{sec:results} we present the derived gas masses and gas mass fractions for the
entire cluster sample, compare these results to other gas mass
fraction work, and discuss the limits this work places on \omegam\ and
plans for future work.

\section{Instrument and Observations}
\label{sec:instrument}
\subsection{An Overview of the OVRO and BIMA 30 GHz SZ Observations}
We wish to take advantage of the characteristically low noise of
interferometer systems while retaining sensitivity on the large
angular scales subtended by clusters.  To do this, we integrated  
centimeter-wave receivers built expressly for this purpose 
into the millimeter-wavelength interferometer systems at the Owens
Valley Radio Observatory (OVRO) Millimeter Array and at the
Berkeley-Illinois-Maryland Association (BIMA) Millimeter Array.  
The angular scale sampled by an
interferometer element is $\theta \sim \lambda/B$, where $B$ is the
projected baseline, or telescope spacing as seen by the source.  
At our $\sim$ 1 centimeter operating wavelength, the compact telescope 
configurations effectively sample the angular scales of clusters 
while the fluxes of any contaminating pointlike sources in the field
are simultaneously monitored
with the longer baselines, so their time variability is not a source
of uncertainty. Operating the millimeter systems at $\sim10$ times
lower frequency than the design frequency also provides for very good
optical performance.  Both arrays allow for the elements to be placed
in a wide variety of configurations.

The receivers, which operate from 26 to 36 GHz, are based on cryogenically-cooled 4-stage HEMT
amplifiers and achieve receiver temperatures of 12-20 K at 28.5 GHz. 
Results from this system are reported in \cite{carlstrom1996,carlstrom2000,grego2000a,holzapfel2000,holzapfel2000b,patel2000,reese2000}.

\subsection{The Interferometric Arrays}
We observed with this system at the OVRO Millimeter
Array in the summers of 1994-1996 and 1998.
At OVRO, the weather was adequate for observing about 80\% of the
time.  
The aperture efficiency at 28.5 GHz, $\simeq$ 0.75,  was measured with
holographic techniques.  The contribution of the antenna to the system
temperature, including spillover, is $\simeq$ 12-15 K, as measured
from sky dips.  The array of six 10.4 meter telescopes (four
telescopes in 1994) is two-dimensional, with baselines
ranging from 14 to 240 meters.   A general description of the OVRO millimeter array is provided in \citep{padin1991}.
The continuum measurements are made with the dual-channel analog correlator, each channel having an input bandwidth of 1 GHz.
In 1994, the SZ observations were made using a single channel centered
at 28.7 GHz. After 1994, the observations were made in single-sideband
mode using two 1 GHz
channels, centered at 28.5 and 30 GHz.  At OVRO's latitude, sufficient
\uv\ coverage can be obtained for sources with declinations greater
than $-10^\circ$ when two or three different telescope configurations
are used.  The primary beams for each telescope are measured holographically
and are quite similar. The full width at half maximum (FWHM) of the
beams differ maximally by five percent, and can be approximated as Gaussian with a
FWHM of 252$''$.

We used the same receiver system at the BIMA Millimeter Array in the
summers of 1996, 1997, and 1998.  In 1996, we used the six receivers
on six of the BIMA telescopes; three additional receivers were
constructed to use a total of nine of the ten BIMA telescopes in the summers of 1997 and 1998.  
At BIMA, the contribution to the system temperature from the antenna is minimal, $\sim$ 6 K.  The aperture
efficiency at 30 GHz with our receivers is $\sim$ 0.70.  The BIMA array is two-dimensional, with baselines ranging from as short as 7.5 meters and as long as 1 kilometer.
  A general description of the BIMA
interferometer is given in \citep{welch1996}.  We operate the hybrid
digital correlator in wideband mode (mode 8 in the notation of \cite{welch1996}) covering 800 MHz with 2-bit sampling.  Adequate \uv\
coverage can be obtained for sources with declinations greater than
about $-10^\circ$ with one or two telescope configurations.
The primary beams for each telescope were measured holographically and
are very similar. The FWHM of the beams differ maximally by $\sim$3\%
and can be approximated by Gaussian with a $396''$ FWHM.

\subsection{Data Reduction and Calibration}
\subsubsection{OVRO Reduction}
\label{sec:OVROredu}

Our observing strategy maximized usable observing time
on the cluster while also providing reliable instrument calibration.
During times the cluster was observable with minimal shadowing, we interleaved
twenty-four minute observations of the cluster with observations of a
nearby bright radio source (the gain calibrator) to monitor the
stability of the interferometer's phase and amplitude response.
The cluster and gain calibrator observations were taken in several short
segments (four and one minute integrations, respectively) to
minimize the effect of short-term instabilities on observing
efficiency. 
Either a planet or a time-stable bright radio source was observed
to provide the absolute flux scale for the measurement.
This flux scale is based on Mars; if Mars was at least $15^\circ$
above the horizon during the cluster observation, it was observed.

The data were edited according to several criteria.  
Data taken with a telescope which was blocked by another telescope
(shadowed) are removed from the data set.  We use a conservative
shadowing limit; data are discarded if the projected baseline is less
than 1.05 times the telescope diameter.  Also removed are data taken
during poor weather as evidenced by poor phase stablity and data affected by anomalous
jumps in the instrument's phase.  Any cluster data not bracketed by calibrator
observations are also removed.

Data calibration proceeds in two steps, gain calibration and absolute flux
calibration.  
A time series of the gain calibrator's amplitude and phase in each
baseline is examined
with the MMA data reduction package \citep{scoville1993}.  The
instrument response during the cluster observations is interpolated from a
fit to this time series.  The amplitude response generally varied 
less than a percent over an observation of many hours.  
The average gains for each
baseline were quite stable from day to day.  In the 1994 and 1995
observing seasons, however, the receivers responded to linearly polarized light.  Since some of
the gain calibrators are linearly polarized at the 5-10\% level, the 
measured amplitude of such calibrators changes with parallactic angle
as well as instrument response.   
Only a few of the cluster observations are significantly
affected, since two different gain calibrators were often used for a single 
cluster and in no case were both noticeably polarized.  For the affected clusters, the average calibrator flux is used and
the amplitude gain is assumed to be constant.  
The instrumental phase response typically only drifts a few tens of degrees over the
course of a cluster observation.

The absolute flux scale is determined relative to Mars.  Mars'
brightness temperature is predicted using a radiative transfer model 
for the whole disk brightness temperature \citep{rudy1987} for each day of
observation.  The intrinsic
uncertainty of this model is expected to be $\sim 2.5\%$ and the
uncertainty from input parameter uncertainties is about 3\%, and so we
estimate the accuracy of our absolute flux scale to be 4\% at 90\% confidence. 
We calculate the brightness
temperature at the center frequency of our observed band; the 
brightness temperature varies less than a percent over our bandpass. 
The solid angle Mars subtends at each observation is determined from the equatorial and polar
diameters of Mars reported in the Astronomical Almanac.

\cite{goldin1997} compared the Rudy model to a thermal model for
Mars, and find even with substantial extrapolations in wavelength, the
two models predict brightness temperatures for Mars consistent with
each other.  We also compare the Mars brightness temperature predicted
by the Rudy model to those derived by \cite{mason1999} based on
absolute flux measurements of Cas A.  They find Mars' brightness
temparature at 32 GHz to be $196.0^{+7.5}_{-7.6}$ K for the May 1998
epoch.   In our observing scheme, we determine the brightness
temperature separately for each day.  The brightness temperature
predicted by the Rudy model at 32 GHz varies from 194 K to 203 K
during the month of May 1998.  The brightness temperature for Mars
varies less than 0.2\% between 28 and 30 GHz.  These comparisons
suggest that our primary calibration is accurate and is consistent
with the primary calibration used by other groups.

The fluxes of a set of primary calibrators were
determined using the predicted Mars flux.  Since the amplitude gain of the
instrument is stable with respect to time and telescope elevation, the
observations of these calibrators and Mars need not be contiguous.  
In the case our primary calibrator is never observable at the same
time as Mars,  we bootstrap the flux from another primary calibrator.
 Over each of the month-long observation seasons, no time variation of the gain calibrator fluxes was evident.

\subsubsection{BIMA Reduction}
\label{sec:BIMAredu}
At BIMA, we use an observing scheme similar to that used at OVRO, interleaving observations of the gain calibrator and cluster.  

The reduction proceeds much like the OVRO reduction with additional
editing and passband calibration.   Spectral channels with low signal-to-noise ratio or with spurious interference are discarded.  Also edited out are shadowed data, data taken with obviously incorrect 
or irregular system temperatures, data taken when the telescope
tracked incorrectly, and data contaminated by local interference.  (These
errors are flagged online at OVRO.)  The spectral response of the
instrument is determined from a passband calibrator, and then the
spectral channels are vector averaged into one wideband channel.  The
gain calibration is then performed.

Absolute flux calibration at BIMA evolved between the 1997 and 1998 seasons.
For the 1997 data, each of the Mars observations were reduced in the method
described above, and the resultant amplitude and phase are SELFCALed.
The amplitude response for each of the 9 telescopes is determined using the 
flux of Mars from the Rudy model. The gains were very stable over the two months
of observing time, with an \rms\ antenna gain variation of 1.2\% for
all telescopes all summer.  With the knowledge that the amplitude
response is steady, in 1998 the gains were derived in the first week of the BIMA observations, and these gains are
applied online.  Mars was subsequently observed to monitor any gain variation.

\subsection{Cluster Selection and Observations}
\label{sec:sample and results}
We observed over 40 clusters with the
centimeter-wave SZ system during the 1994-1998 observing seasons.
Only some of these clusters were observed for a significant amount of
time; some observations were intended to survey for point sources and to
define a sample for future work.  To date, over 25 clusters have been detected significantly; analysis of a sample of 18  are presented here.

The cluster targets were selected from a flux-limited, homogeneous sample of clusters \citep{stocke1991,gioia1994,nichol1997} identified from
the {\em Einstein} Extended Medium Sensitivity Survey (EMSS)
\citep{gioia1990} and from two flux-limited samples (XBACS, \cite{ebeling1996}; BCS, \cite{ebeling1998}) from the {\em ROSAT} All-Sky Survey,
as well as public {\em ROSAT} data.    
Identifying clusters based on X-ray emission rather than galaxy
surface-density enhancements ameliorates the problems of false
detections due to chance superpositions and of missed clusters due to
smaller-than-average backgrounds.  

We selected massive clusters for our sample.  X-ray studies of
clusters in \cite{djf95} and \cite{mohr1999a} indicate that the gas mass
fraction near the virial radius increases as cluster mass increases, but that above $\sim 5$ keV the gas mass fraction at the virial
radius is constant.  At the initiation of this work, X-ray temperatures
were not widely available for distant clusters.  We chose instead to
select clusters on the basis of luminosity.  We expect a cluster's X-ray
luminosity to be a better predictor of mass than X-ray surface
brightness, as it will be less sensitive to projection effects and
contamination by cooling flows and dynamical activity in the ICM. 
Although cooling flows have been observed to contribute as much as
70\% of a cluster's luminosity, typically they only contribute 10-30\%
\citep{peres1998}.  Subsequent X-ray spectral measurements confirm that the clusters in
this sample all have emission-weighted temperatures greater than 5
keV and therefore qualify as massive clusters for our purposes.

Our SZ observing scheme requires the clusters to be at
declination greater than $-10^\circ$. The apparent size of the cluster
must also be small enough so that the angular size is comparable to the spatial
frequencies the interferometer samples; this is generally satisfied if
the cluster redshift is greater than about 0.14. 
For the initial cluster observations, we did not pursue observations towards
cluster fields which hosted point sources with flux densities greater than
$\sim$10 mJy; fewer than $\sim$15\% of cluster fields had such point
 sources.  We have since confirmed that we can reliably remove such
 point sources from the data, and we are pursuing observations towards
these fields. 
 
It is possible that selecting against clusters with strong point
sources may introduce a bias.  This bias would be redshift
dependent because, while the SZ effect magnitude is not diminished by
distance, the flux of a point source associated with the cluster 
is.  Clusters with radio-loud central galaxies will be less likely 
to be dropped because of point source contamination if they are distant.  \cite{peres1998} study a sample of 55 nearby X-ray clusters, 40\% with inferred
cooling flow mass deposit rates of over 100 $M_\odot$/yr.  Forty-one of
these clusters have radio detections or upper limits at 1.4 or 5 GHz,
and 33 of these have detected radio flux.  \cite{peres1998} cross-correlate the radio data and find only a weak correlation 
between the radio power of the brightest cluster galaxy and the
strength of the cooling flow.  They do find that the largest cooling
flows have the strongest radio fluxes, though.  By
selecting against clusters with very strong radio emission, we may be
removing from our sample clusters which have not undergone recent
mergers strong enough to disturb a cooling flow.  Again, we expect
this effect to be small, if present, as 85\% of clusters were kept in the sample.

\begin{figure}
\figurenum{1}
\centerline{\hbox{
\psfig{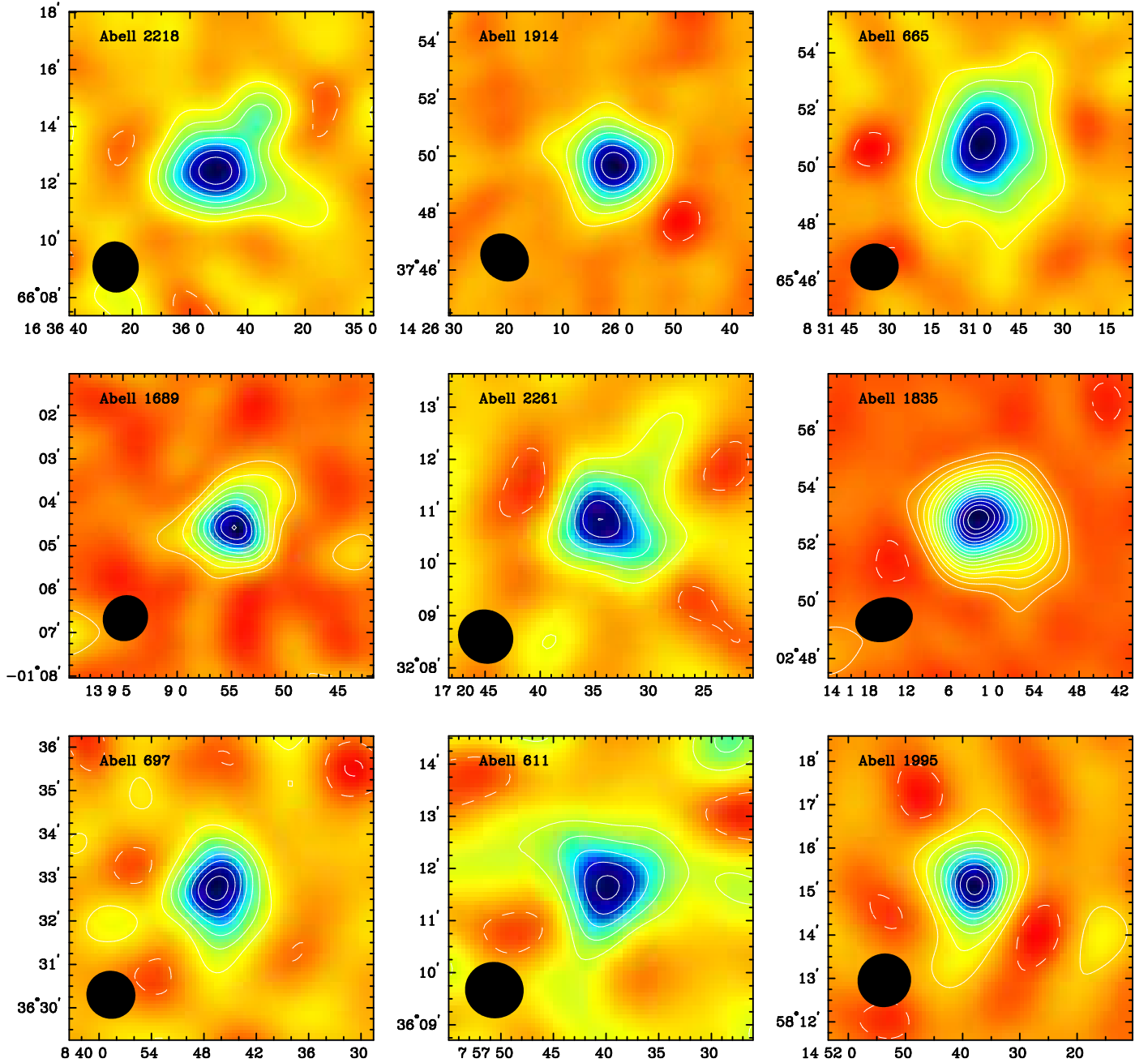}}}
\caption{The CLEANed images of the clusters in the sample are
presented in order of increasing redshift. The images are plotted with
1.5 $\sigma$ contours, and the restoring beam is shown in the bottom left
corner of each image.\label{fig:clustimgs}}
\end{figure}

The 18 clusters in our sample are listed in Table~\ref{table:sample},
along with the published redshifts and $T_e$ we used in the $f_g$
analysis, and the X-ray luminosities.
SZ images of the clusters are presented in Figure~\ref{fig:clustimgs}, 
ordered by redshift.  We note that the quality of the detection 
reflects the \rms\ sensitivity of the observation and the intrinsic
strength of the SZ effect and not the cluster's distance.  
A Gaussian taper is applied
to the $u-v$ data to emphasize the structure on large scales.  This
taper depends on the range of \uv\ radii in each cluster's data set;
the tapers are generally 0.9 to 1.2 k$\lambda$.  Higher
resolution images can be made from these data in order to emphasize
smaller cluster structures,\eg, \cite{carlstrom1996}.  Because the
primary beam for the BIMA system is considerably larger than that for
the OVRO system (396$''$ and 252$''$, respectively), the images
produced from BIMA data include information on the decrement at larger scales.  The images are plotted in contours of
1.5 $\sigma$, and the restoring beam is shown in the bottom left
corner of each image.  

The interferometric SZ data is necessarily spatially filtered; the
visibility function will not be measured at every spatial
frequency. The images are used to indicate the signal-to-noise ratio of the
cluster detections, but we do not fit models to the images.

\section{SZ Gas Mass and Gas Mass Fraction Measurement Methods}
\label{sec:fgmethods}
\subsection{Model}
  We compare a model to the data in the spatial frequency domain,
where the noise characteristics and the spatial filtering of the
interferometer are well-understood.

We fit to a \be-model \citep{cavaliere1976,cavaliere1978}, which has been widely used to fit the density
and temperature profiles of cluster galaxies and the ICM.  We make the
simplifying assumptions that the cluster gas is isothermal and the
density distribution is spherically symmetric.  We consider the effects of these
simplifications on our results in Section~\ref{sec:szfgsyserr}.

In this model, the electron number density as a function of radius, $r$, takes the form:
\begin{equation} n_e(r) = n_{e\circ} \left ( 1 + {r^2 \over r_c^2} \right)^{-3 \beta/2}, \label{eq:density}\end{equation} 
where $r_c$, the core radius, and \be\ are fit parameters, and 
$n_{e\circ}$ is the central electron number density.   For isothermal gas
with temperature $T_e$, Equation~\ref{eq:density} predicts the following  two-dimensional SZ temperature decrement:
\begin{equation} {\Delta T}\left ( \theta \right ) = \Delta T(0) \left( 1 +  {\theta^2 \over \theta_c^2} \right)^{{1\over 2} -{3\beta \over 2}},\label{eq:DT/T}
\end{equation}
where $\theta = r/D_A $, $D_A$ is the angular diameter distance,
$\theta_c = r_c / D_A$, and 
$\Delta T(0)$ is the temperature decrement at zero projected radius.
The central electron density can therefore be recovered from this relation:
\begin{equation}n_{e\circ} = {-\Delta T(0)\over T_{CMB}}{m_e c^2 \over 2
k \sigma_T} {1 \over T_e}\left(D_A \int\limits_{-\infty}^{+\infty} \left(1 + \left({\theta \over
\theta_c}\right)^2\right)^{-3\beta/2} d\theta\right)^{-1}\label{eq:ne0}\end{equation}
where the integral, $dl$, is along the line of sight.  The mean molecular weight is assumed to be constant
througout the gas, so the electron number density, $n_e$, should
trace the gas density.  

\subsection{Fitting Procedure}
\label{sec:modproc}

The change in
spectral intensity of the CMB due to the Sunyaev-Zel'dovich effect is calculated for the Rayleigh-Jeans approximation (\cf\ \cite{rephaeli1995,challinor1998}): 
\begin{equation}\label{eqn:dT(x)}{\left .{\Delta T_{SZ}\over
T_{CMB}}\right |_{RJ} = {y x^2 e^x \over (e^x -1)^2}[ x coth(x/2) - 4 + \theta_e f(x)],}\end{equation}
 where $x = h\nu /  kT_{CMB}$ and $\theta_e = kT_e / m_e
 c^2$. We adopt the COBE FIRAS value of $T_{CMB} = 2.728$ K \citep{fixsen1996}. The last term, $\theta_e f(x)$, corrects for relativistic
 effects.  
  At 28.5 GHz, $\Delta T_{SZ}/ T_{CMB} = -1.92\ y$ in the non-relativistic Rayleigh-Jeans approximation.   Including the relativistic correction for a temperature typical of massive clusters, $kT_e = 7$ keV, $\Delta T_{SZ} / T_{CMB} = -1.88\ y$.  

 The data are
components of the Fourier transform of the sky brightness
distribution, \ie,  a measured amplitude and phase for each
two-dimensional spatial frequency, or \uv\ pair, sampled.  
The model is constructed in image
space by filling out a regular grid with the SZ model (Equation~\ref{eq:DT/T})
multiplied by the primary beam.  This SZ image is fast Fourier
transformed and the model is interpolated to the $u$-$v$ data points
to compare with the data using the $\chi^2$ statistic. The cluster
center, $\beta$, $\theta_c$, and $\Delta T(0)$ are allowed to float to
find the minimum $\chi^2$ using a downhill simplex (Press et
al. 1992).  The position
and flux density of any radio-bright point sources are also fitted.
Since the primary beam attenuation at any given point differs
between the OVRO and BIMA datasets, and the intrinsic point source
flux can vary with time, the point source fluxes for each dataset are allowed to vary individually.

The fits are performed jointly on all datasets for a given cluster. 
The shortest telescope spacing corresponds to the shadowing limit; for OVRO data
this limit is $1\ k\lambda$, for BIMA data this is $0.58\ k\lambda$.  
We use the holographically determined primary beam models 
when modeling the data, and the entire datasets are 
used to do the analysis. 

\subsection{Constraints on Fit Parameters}
\label{sec:modconstraints}
The cluster's centroid position and the point source fluxes and
positions are well constrained by the data.  
The fitting program consistently obtains the same values for the centroid
positions.   The initial guesses
for the point source parameters are made using DIFMAP (\cite{pearson1994}), an interactive mapping program, to inspect the
high spatial frequency ($|u^2 + v^2| > 2.0$ k$\lambda$) data in which
the SZ effect contributes very little
signal.  The uncertainty introduced by point sources into the ICM parameters is discussed in Section~\ref{sec:sysradio}.

\begin{figure}
\figurenum{2}
\epsscale{0.85}
\plotone{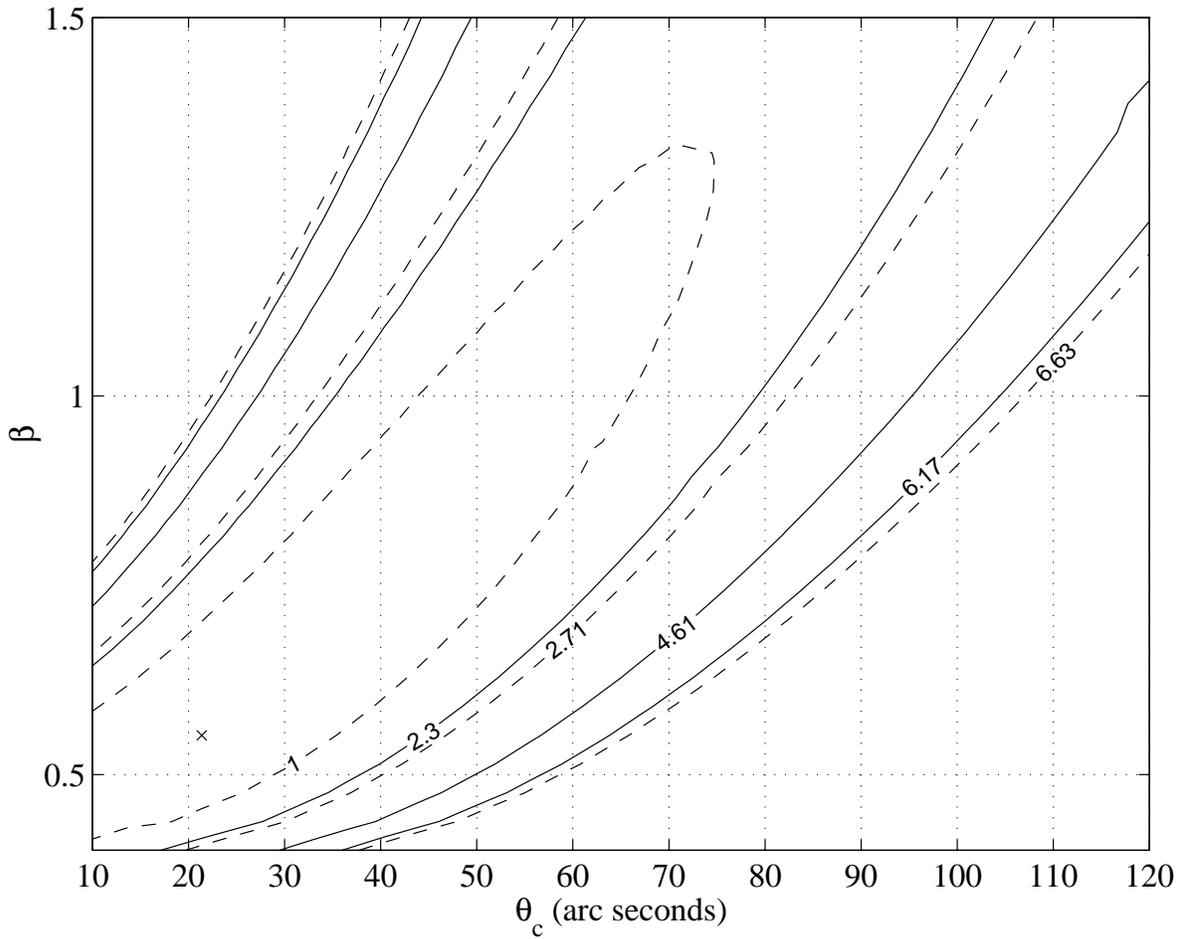}
\caption{Confidence intervals for \be\ and \rc\ from a \be-model fit
to the Abell 1995 data.   At each point in
the plot, the central decrement was allowed to assume its best fit
value. The solid contours are marked for \dchi\ = 2.3, 4.61, and 6.17 which indicate 68.3\%, 90.0\%, and 95.4\% confidence, respectively, for the two-parameter fit.  The dashed lines indicate \dchi~=1.0, 2.71, and 6.63.  The projection onto the \be\ or \rc\ axis of the interval contained by these contours indicate the 68.3\%, 90\% and 99\% confidence interval on the single parameter.}
\label{fig:beta_rca1995}
\end{figure}

The cluster centroid and point source fluxes and positions are fixed to their best-fit
values while the cluster shape parameters are fitted.  We found no
appreciable variation of best fit centroid position with shape
parameters.  To illustrate the constraints these data place 
on \be\ and \rc\, a grid search is
performed over these parameters with \dt\ allowed to assume its best fit value at each grid point. In Figure~\ref{fig:beta_rca1995}, we show the
confidence intervals for \be\ and \rc\ for a representative cluster,
Abell 1995. The solid contours indicate \dchi\  = 2.3, 4.61, and 6.17
which enclose regions corresponding to 68.3\%, 90.0\%, and 95.4\% confidence, respectively, for the two-parameter fit.  The
projection of the dashed lines, \dchi\ $=1.0$, 2.71, and 6.63,
indicate the 68.3\%, 90\% and 99\% confidence interval on the single
parameter.  At each (\be, \rc) point, the width of the 68\% confidence
interval for \dt\ is about 10-15\% of the best fit \dt\ value.  
In \cite{patel2000}, we fit the \rosat\ HRI data for this cluster,
and find the fit values to be consistent with the SZ effect values.

From the figure it is clear that \be\ and \rc\ are strongly correlated
and the fit parameters \be\ and \rc\ are not well-constrained
individually by the SZ effect data.

\subsection{Gas Mass Fraction Measurements}
\label{subsec:gasmassfraction}
The number of electrons in a given volume can be calculated by
 integrating Equation~\ref{eq:density}.  To recover the ICM mass, we
 multiply by the proton mass and the nucleon/electron ratio of 1.16.
 To extract the central electron density, $n_{e\circ}$, for a given
 set of model parameters and measured electron temperature, we perform
 the integral in Equation~\ref{eq:ne0}.
Formally, this integral extends from the observer along
the line of sight through the cluster infinitely; in practice, a
cutoff radius of 8-10 cluster core radii is used.

We note that although the fit parameters \be, \rc\ and \dt\ are not
 constrained strongly individually, the combination of these three
 parameters does constrain the gas mass quite well.  This follows from
 the fact that the SZ effect is, under the isothermal assumption, a
 direct measure of the gas mass on the scales to which our
 observations are sensitive.  We present gas masses for the 18 clusters in our sample in
 Table~\ref{table:fgcirc}.  To convert angular sizes to lengths,  we have
 assumed $h=0.7$, \omegam = 0.3, \omegal=0.7.

The distribution of the cluster's total mass, mainly comprised of
dark matter, can be inferred from the modeled gas pressure
distribution, since the temperature of the gas and its  spatial 
distribution are
constrained by the cluster's gravitational potential.  We make the 
assumption that the gas is in hydrostatic equilibrium in this
potential and that bulk flows and other non-thermal processes do not 
contribute significantly to the gas pressure.  
Under the assumptions of spherical symmetry and isothermal gas, the total
mass of a cluster within radius $r$, is
\begin{equation}M(r) = {3 k T_e \beta \over G \mu m_p} {r^3\over
r_c^2 + r^2}\label{eq:totalmass}\end{equation}
where $\mu m_p$ is the mean molecular weight
of the gas.  To calculate $\mu$, we assume the gas has solar metallicity as measured by 
\cite{anders1989} and that $\mu$ is constant throughout the
gas.  The value of $\mu$ will change 3-4 \% depending on the solar
metallicity measurements one adopts;  the metallicity in clusters is
not well known, however, and although an incorrect choice for $\mu$  will
introduce a systematic error, it will be much smaller than the statistical
errors involved.  Note from Equation~\ref{eq:totalmass} that the total
mass depends only on the shape of the gas distribution, and 
is independent of the value of the central gas density, and therefore
of the uncertainties in \dt. 
  Using the
derived shape parameters, \be, \rc, and the measured gas temperature,
  we derive the total mass, denoting it the ``HSE mass''.

To measure the quantities of interest and their associated
uncertainties,  we determine an appropriate range 
\be, \rc, and \dt\ for each cluster with a coarse grid, and then construct a finer grid near the best fit parameter values.  
The cluster's gas mass, HSE mass, gas mass fraction, and 
$\chi^2$ statistic are derived at each grid point.  The 68\%
confidence interval on each quantity is determined from the range contained in the $\chi^2$(best fit) - $\chi^2$
= \dchi$<$ 1.0 volume of the parameter grid.
We prefer to measure the masses and mass fractions in the largest volume permitted by our method, since the fair sample assumption is best at large radii.  
The largest spatial scale on which we can constrain the model depends on the
\uv\ points at which we significantly detect signal.
To determine this scale,  we calculate the statistical uncertainties in the
$f_g$ measurement due to the shape parameter uncertainties for a
number of radii from 10$''$ to 150$''$.  We find we best constrain
$f_g$ when it is calculated within a radius of around 65$''$ (see \cite{grego1999a}).
The gas masses and gas mass fractions within a 65$''$ radius along with their associated 68\%
confidence intervals are presented in Table~\ref{table:fgcirc}.  The gas mass and $f_g$ results depend on the assumed cosmology through
the angular diameter distance, \da.  For the gas mass fractions reported, 
we use $\Omega_M =
0.3, \Omega_\Lambda = 0.7$.

The SZ gas mass is inversely proportional to the assumed electron temperature:
$M_{gas}{\rm (SZ)}\propto 1/T_e$ and the HSE total mass measurement is directly proportional to $T_e$: $M_{HSE}\propto
T_e$.  The gas mass fraction then is quite sensitive to temperature:
$f_g \propto 1/T_e^2$. The uncertainties from the temperature
measurement are of the same order as the statistical uncertainties
from the SZ model fitting at the lower redshifts, and dominate the SZ uncertainties for the
most distant clusters.

\subsection{Systematic Effects}
\label{sec:szfgsyserr}
\subsubsection{Emission-Weighted Temperature}
When available, we have used emission-weighted temperatures which were
examined and corrected for the presence of cooling flows.  The central surface
brightness excess exhibited by many clusters is interpreted as emission
from centrally-concentrated dense gas, \eg, \cite{fabian1994}, the cooling time of which is shorter than
the Hubble time. Such cooling flows can bias the emission-weighted
temperatures lower than the density-weighted or virial temperature of
the cluster.  \cite{allen1998} find that
modeling clusters with a cooling flow spectral component in addition to the thermal
component significantly reduces the scatter in the luminosity-temperature
relation.  We have used these cooling flow-corrected temperatures where
available.  

The emission-weighted ICM temperatures we have adopted from the
literature may also have errors due to contamination from other sources 
in the field.  The \asca\ observatory was the source for most of the
published ICM temperatures we use in this work.  As its half power
diameter is $\sim 3'$, it is nearly impossible to remove the effects of point
sources on spectra of distant clusters obtained with \asca. Since the measurement
is so strongly dependent on an accurate measurement of $T_e$, this is likely to be
the largest source of systematic uncertainty.  Fortunately, many of
the clusters in our sample are scheduled to be observed with the
{\em Chandra} and {\em XMM} observatories, which will be better able 
to distinguish ICM emission from point source emission and toconstrain the ICM temperatures.

\subsubsection{Radio Point Sources}
\label{sec:sysradio}
We detect radio-bright point sources in about half of the observed
clusters.  The point sources with fluxes exceeding three times the
\rms\ of the high resolution ($\gtrsim 2000 \ \lambda$) maps can be reliably identified from the
SZ data. We estimate the maximum effect of undetected point sources by adding
an on-center point source to a representative cluster data set and
fitting this new data set not accounting for the added point source.
We place a 3
$\sigma$ point source at the cluster center where typical \rms\
sensitivities in the OVRO and BIMA high resolution maps are roughly 61
$\mu$Jy and 163 $\mu$Jy respectively.  This point source causes the
magnitude of the decrement to be underestimated (and thus the gas mass
fraction too) by 15\% for OVRO data and 20\% for BIMA data.  Such a
point source at the cluster center is highly unlikely but places
limits on the maximum effect from undetected point sources.


\subsubsection{Departures from an Isothermal, Spherical ICM}
\label{sec:tempgradnonsphere}

Our assumptions that the intra-cluster medium is isothermal and
spherical are at some level approximations.  \cite{markevitch1998} report 
moderate temperature gradients in a sample of 30 nearby clusters,
although in a similar analysis, \cite{irwin1999} do not find such
structure.
Neglecting to account for existing temperature gradients in the ICM 
may systematically affect the gas and HSE masses.  If such a gradient
is present, 
the true temperature in the central region may be higher than the 
emission-weighted temperature we use, and the fitted shape
parameters from the isothermal SZ analysis may no 
longer accurately describe the density distribution.  
As yet, there are no strong observational constraints on 
temperature structure in clusters beyond z = 0.1, as
there have been no suitable telescope facilities for the task.
We anticipate that the {\em Chandra} and {\em XMM} X-ray observatories will
greatly improve this situation.

Our observation scheme provides information on the two-dimensional decrement, and we observe that the clusters are not strictly
spherical.  For this sample of clusters, we find the mean of the
best-fit axis ratios to be 0.89 $\pm$ 0.12.  In previous work (\cite{grego2000a,grego1999a}), we relaxed this assumption and permitted the density 
distribution to be ellipsoidally symmetric, but the unknown
orientation and three-dimensional geometry introduce a large
uncertainty in the HSE mass.  For a sufficiently large 
sample chosen without orientation bias, deviations from spherical
symmetry will not strongly affect the results.  As a point of
comparison, the effects of cluster
shapes on determinations of the cluster size
have been investigated in \cite{sulkanen1999}.  He calculates Hubble's
constant in a sample of simulated tri-axial clusters by comparing
their predicted SZ and X-ray images.  The X-ray flux from the cluster 
at any point in the sky is proportional
to $\int n_e^2 dl$, integrated along the line of sight through the cluster, while the SZ effect is
proportional to $\int n_e dl$, so the two can be compared to derive
the size scale of the cluster; when this size scale is compared to the
apparent size on the sky, the cluster's distance and hence $H_\circ$
can be inferred.  Sulkanen finds that when the images
are fit by an spherical beta model with a core radius equal to 
the arithmetic mean of the two core radii from an elliptical fit, the recovered $H_\circ$ for a sample of 25 clusters 
is unbiased.

In an ongoing analysis of an
ensemble of hydrodynamical cluster simulations, we also find that we
do not introduce serious error with these assumptions.    These
simulated clusters are produced within both low and high
\omegam\ cosmological models, and the temperature and density
structure is appropriate for cluster populations
experiencing merging similar to that observed at redshifts z$\le$ 0.1
\citep{mohr1995}.  We produce mock BIMA observations of simulated
clusters at the redshifts z=0.2 and z=0.6.  Isothermal,
spherical \be-model analyses of these SZE observations
produce unbiased estimates of the ICM mass enclosed within
the radius $r_{2000}$, which roughly corresponds to the scales
measured in this experiment \citep{mohr2000}.
Should temperature and density structure 
in distant clusters be similar to that in the local sample, it should
not be a significant source of systematic uncertainty or bias in our measurements.

\subsubsection{Validity of the HSE Approximation and Non-thermal Pressure Support}
Our method of measuring the total mass assumes the ICM is in
hydrostatic equilibrium in the cluster potential and supported only by
thermal pressure.  One test of this assumption is to
compare the HSE-derived total cluster mass to the total mass derived from gravitational lensing models.
Some mass comparisons in the literature (\cite{miralda1995,loeb1994,wu1997}) have suggested that the HSE method may systematically underpredict 
the cluster's total mass by a factor of 2-3, compared to a strong
gravitational lensing analysis. 
Suggested explanations for this discrepancy include elongation of the
cluster along the line of sight and temperature structure in the ICM,
which we discuss in Sections~\ref{sec:tempgradnonsphere}, and non-thermal pressure support of the gas in the cluster core.

Further work has suggested that the details of the analysis can have a
significant effect, and may resolve the discrepancy.  A weak lensing analysis was performed by \cite{squires1997} on the cluster Abell 2218, which appears to have discrepant masses
in each of the three analyses above.  This analysis, at larger radius
than the strong lensing analyses, show the two methods predict
masses which are consistent within the experimental uncertainties.  In
an examination of a sample of 13 clusters, \cite{allen1998} finds
that the lensing and HSE masses agree for clusters which appear to
have a strong central cooling flow, when the cooling flow is taken
into account;  in these clusters, the X-ray and
lensing core radii are consistent with each other, and the mass
agreements suggest HSE is a reasonable approximation.  For clusters without strong cooling flows, the X-ray
core radii are generally larger than the lensing radii, and offsets
between the centers of the distributions are observed, suggesting that
HSE is not appropriate for the cluster cores.  Outside the cluster
core (at radii $\ge \sim$ 400 kpc, weak lensing and X-ray masses are consistent with each other
both for cooling flow and non-cooling flow clusters. \cite{lewis1999} compare X-ray HSE masses to
the dynamical masses calculated from the galaxy velocity dispersions,
and find the average M$_{dyn}$/M$_{HSE}$ to be 1.04$\pm$0.07,
which also suggests the HSE method does not introduce a systematic bias.

Possible sources of non-thermal pressure in the ICM are bulk flows
and magnetic fields in the gas.  Intracluster magnetic fields are typically a few $\mu$G
(\cite{kim1990,taylor1993}), an order of magnitude
smaller than the level at which the fields would contribute
significantly to the dynamics of cluster gas, although stronger
fields, $\sim 10-$100 $\mu$G, have been measured in a few clusters
(\cite{taylor1993}).  There is some evidence for the persistence of 
bulk flows in clusters undergoing merger events (\cite{bliton1998}).  It remains to further investigation how significant a role
these effects play in the physics of
cluster gas, but currently there is no evidence to suggest a
significant systematic error in the HSE method.


\subsubsection{Cosmic Microwave Background Anisotropies}

The SZE observations are also sensitive to intrinsic and secondary
anisotropy in the Cosmic Microwave Background (CMB) radiation. The
theoretical expected and observed level of CMB anisotropy at the
angular scales corresponding to those used for the SZE measurements
presented here is small and safely ignored. The contribution of
primary anisotropy for a window function appropriate for our shortest
baselines, for which the contribution would be strongest, has been
calculated by \cite{holzapfel2000}. For a flat universe, as indicated
by recent CMB observations \citep{miller1999,debernardis00,hanany00} the \rms\ temperature fluctuations
contributed by primary CMB anisotropy within our maps should be of
order $2\mu$K or less.

At the angular scales of our SZE measurements, secondary CMB
anisotropy is expected to be stronger than intrinsic
anisotropy. \cite{holzapfel2000} have tabulated the expected range
of the magnitude of the temperature anisotropy due to the Visniac
Effect and inhomogenous reionization. The upper range for the \rms\
temperature fluctuations at angular scales appropriate for the SZE
measurements is only 5.6 $\mu$K and 3.9 $\mu$K, respectively. Added in
quadrature, this gives an expected upper limit to the \rms\ temperature
fluctuations in our SZE observations of only 6.8 $\mu$K. If present,
this signal would contribute to the SZE maps as noise (\ie, would not
lead to a bias in our SZE measurements).  This level is much smaller
than the noise level obtained in our SZE observations.

The dominant contribution to secondary anisotropy at
the relevant angular scales is likely to be the SZE
from undetected low mass clusters. \cite{holder99a}
estimate \rms\ temperature fluctuations of order 2 $\mu$K to 12 $\mu$K
for the range of models they consider. Again, this level is
small compared to our uncertainties, although approaching
the noise level in our deepest fields. 

It is unnecessary to depend on theoretical estimates of contributions
from CMB anisotropy as we have direct measurements of `blank' fields
obtained with the same instrument as for the SZE observations
\citep{holzapfel2000}. The \rms\ level obtained
on the deepest fields ranges from 16 $\mu$K to 20 $\mu$K, just slightly
above that expected from the instrumental noise. The most likely level
of anisotropy, including undetected point sources, derived from the
blank field data is 12 $\mu$K and the 95\% confidence level upper
limit is 19 $\mu$K.

We conclude that temperature fluctuations due to primary and secondary
CMB anisotropy should have a negligible effect on the results derived
from the SZE measurements reported here.

\section{Results and Discussion}
\label{sec:results}
\subsection{Gas Mass Fractions}
As we discussed in Section~\ref{subsec:gasmassfraction}, we measure the gas mass fraction within a fixed angular radius, which
results in the measurements being made at different physical
scales for clusters at different distances.  To compare the
gas mass fractions of different clusters, and to derive a result useful
for cosmological tests, we scale the results for each cluster to a fiducial
radius.    An analytical expression for the variation in $f_g$ with
radius  is suggested by \cite{evrard1997},  based on results in \cite{EMN96} and found to be consistent with the $f_g$ variation
reported in the \cite{djf95} sample.
We use a modified version of this to extrapolate the
gas mass fractions we measure at 65$''$ to the gas mass fraction
expected at $r_{500}$, the radius at which the cluster's total mass
density is 500 times the critical mass density, where the cluster's
baryon fraction should closely reflect the universal value.  
The physical radius at
which the overdensity is 500 depends on the cluster's temperature (a mass
indicator), and also its redshift, since the critical
density will change with $z$.
  The scaling expression is as follows:
\begin{equation}{f_{g}(r_{500}(T_e)) = f_{g}(r_X)\left ({r_{500}(T_e) \over r_X} \right )^{\eta}},\label{eq:fextrap} \end{equation}
where $\eta$ = 0.17, $f_{g}(r_{500}(T_e))$ is the gas mass fraction within $r_{500}$,
and $r_X$ is the radius within which the gas mass fraction is measured.  We
modify Evrard's expression for $r_{500}$, derived for low redshift
clusters, to include the change in the value of $\rho_c$ with
redshift; $\rho_c(z) = \rho_c(z=0)(H/H_\circ)^2$, where 
$H^2=H_\circ^2[(1+z)^3\Omega_M+(1+z)^2(1-\Omega_M-\Omega_\Lambda) +
\Omega_\Lambda]$ (\cf\ Peebles 1993, Eqn.\ 13.3):
\begin{equation}{r_{500}(T_e) = (1.24 \pm 0.09) \left ( {T_e \over 10\
{\rm keV} (H/H_\circ)^2} \right ) ^{1/2} h^{-1} {\rm  Mpc}}.\label{eq:r500} \end{equation}

The gas mass fractions within $r_{500}$ as derived by this relation are presented in Table~\ref{table:fgcirc}.
Figures~\ref{fig:circ_fgvz,T}a. and b. show the gas mass fractions at
$r_{500}$ as a function of $T_e$ and redshift.  We see no correlation
of gas mass fraction with temperature.  We see no significant
evolution of $f_g$ with redshift.  Since $f_g$ depends on the cluster
distance, $f_g \propto$ D$_A$, and therefore the chosen cosmology,
measurement of the gas mass fraction over a range of redshifts could be
used in principle to constrain cosmological models.

We calculate the mean gas mass fraction for the entire cluster sample,
 and derive the 68\%
confidence interval from the \dchi\ statistic to a constant-value fit.  
For the entire sample, assuming \omegam=0.3, \omegal=0.7, we find the
 mean gas mass fraction to be $0.081^{+0.009}_{-0.011}h^{-1}_{100}$.
We also calculate the
mean and uncertainty for $f_g$ in the full sample, using two alternative cosmologies,
(\omegam=0.3, \omegal=0.0) and
(\omegam=1.0, \omegal=0.0), to
 calculate the distances and scaling relation. In Table~\ref{table:results}, we report these and the associated reduced
chi-squared ($\chi^2_{red}$) statistics, which range from 1.021 to 1.056 for the full sample fits.
 The $\chi^2_{red}$ values do not
 differ significantly enough to discriminate between cosmologies, and it is clear that currently the uncertainties are
too large for a cosmological test via geometry.

We also calculate the mean $f_g$ in a homogeneous subsample of five
clusters.  These clusters are the five most luminous clusters in the
flux-limited EMSS cluster sample with $z > 0.26$ and declination $ > -10^{\circ}$:
MS0451.6-0305, MS1137.5+6625, CL0016+16, MS1358.4+6245, \&
MS1054.4-0321.  For our standard cosmology, we find the mean in this
sample to be $0.089^{+0.018}_{-0.019}h^{-1}_{100}$.  In all three cosmologies, 
the gas mass fraction in the
homogenous sample is consistent with the full sample value.

We compare these SZ-derived gas mass fractions to other SZ-derived $f_g$ measurements.  Recent cluster gas mass fraction measurements from SZ effect observations are
presented in \cite{myers1997}.  In this work, the integrated SZ effect is measured using a single radio dish operating at centimeter
wavelengths.  The integrated SZ effect is used to normalize a model
for the gas density from
published X-ray analyses, and this gas mass is compared to the published total masses to determine the gas mass fraction.  For three nearby clusters, A2142, A2256 and the Coma
cluster, \cite{myers1997} find a gas mass fraction
of $(0.061\pm0.011) h_{100}^{-1}$ at radii of 1-1.5 $h_{100}^{-1}$ Mpc; for the
cluster Abell 478, they report a gas mass fraction of $(0.16\pm0.014)h_{100}^{-1}$.

\subsection{Comparison of SZ and X-ray Results}
\label{sec:szxray}
 Gas mass fractions derived from X-ray images for a large,
homogeneous, nearby sample of clusters are presented in \cite{mohr1999a}.  For a subsample of 28 clusters with $T_e > 5$ keV, they find the mean
gas mass fraction within $r_{500}$ to be $(0.0749\pm 0.0021)h_{100}^{-3/2}$ at 90\% confidence.
The gas mass fractions derived from SZ measurements depend differently
on the cosmology assumed than those derived from X-ray images, and
this should be noted when comparing the results.  

Qualitatively, though, the comparison does not suggest any large
systematic offsets.  This is a significant result, because a large
clumping factor, $c=<n_e^2>^{1/2}/<n_e>$, has been suggested as an
explanation for the high gas mass fractions in clusters \citep{white1993,evrard1997}.  A cluster with clumping factor $c$ would only
require 1/$c$ as much gas mass to produce the observed emission, and
so the SZ and X-ray gas mass fraction measurements would differ by a
factor of $\sim c$.
 \subsection{Comparison of Baryon Fraction with $\Omega_B$}

  The relative abundance of deuterium and hydrogen provides a
particularly strong constraint on the baryonic matter density \citep{copischrammturner}.  A firm upper limit to $\Omega_B$ is set by the presence of deuterium in the local interstellar medium.  This constrains the value of
$\Omega_B$ to be less than $0.031 h_{100}^{-2}$ \citep{linsky1995}.
Measurements of the D/H ratio in metal-poor Lyman-$\alpha$ absorption
line systems in high-redshift quasars put an even more stringent
constraint on the baryonic mass density. For this analysis, we adopt the
published value at 95\% confidence from the \cite{burles1998}
absorption line anaylsis, $\Omega_B=(0.019 \pm 0.002)h_{100}^{-2}$.
 
\begin{figure}
\figurenum{4}
\epsscale{0.85}
\plotone{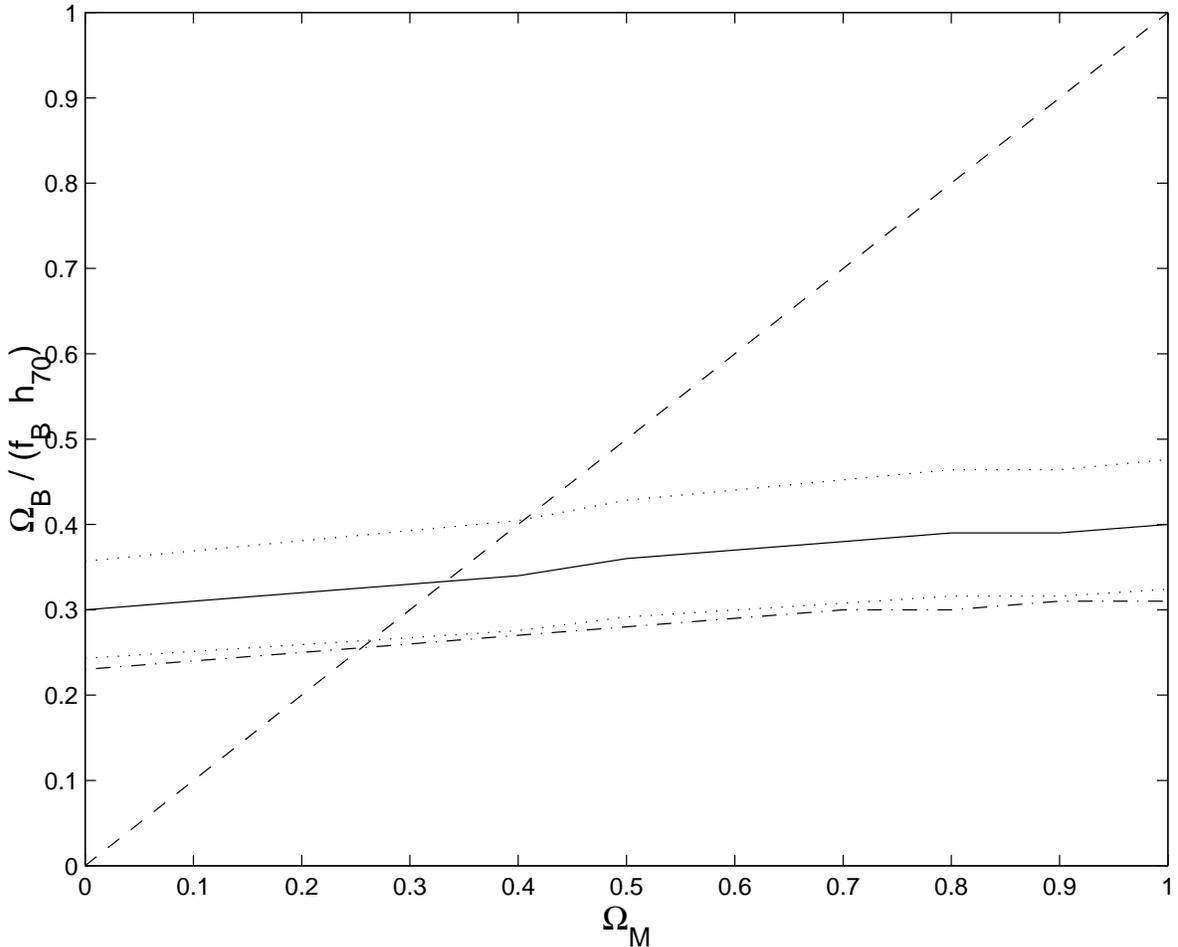}
\caption{Upper limit on the total matter density, \omegam $\le
\Omega_B/(f_B h_{70})$ (full line) and its associated 68\% confidence
region (dotted lines), as a function of cosmology (\omegal $\equiv$ 1
- \omegam).  The baryon fraction is a function of cosmology through
the angular diameter distance relation as well as the scaling relation
in Equation~\ref{eq:fextrap}.  The intercept between the upper dotted
line and the dashed line [\omegam = $\Omega_B/(f_B h_{70})]$ gives the upper limit to \omegam\ at 68\%
confidence.  The dot-dashed line shows the total matter density when
the baryon fraction includes an estimate of the contribution from
baryons in galaxies
and those lost during cluster formation.  The intercept of the dot-dashed line
and the dashed line gives our best estimate of \omegam\ ($\sim$0.25) assuming a flat
universe with $h$=0.7.\label{fig:fitomega}}

\end{figure}

We can 
use the gas mass fractions to
find the value of \omegam\ in a self-consistent manner. In
Figure~\ref{fig:fitomega}, we show the value of \omegam\ implied by
the measured gas mass fractions when we assume a flat universe
(\omegal $\equiv$ 1- \omegam) and $h=0.7$ to calcluate the angular
diameter distance and $r_{500}$ scaling factor from Equation~\ref{eq:fextrap}: \omegam\ $\le
\Omega_B/f_B/h_{70}$ .  The upper limit to
\omegam\ and its associated 68\% confidence interval is shown as a function of \omegam.  The measured gas mass
fractions are consistent with a flat universe and $h=0.7$ when \omegam\ 
is less than
0.40, at 68\% confidence.  For our measurements to be consistent with
\omegam\ = 1.0 in a flat universe, the Hubble constant must be very
low, $h$ less than $\sim$ 0.30.

For a more realistic
estimate, we could include the baryon contribution from galaxies, and
attempt to account for the overall dimunition of the baryon fraction
in clusters with respect to the universal value, since some baryons
are expected to not become bound to the cluster.  Following \cite{white1993}, we estimate the galaxy mass to be a fixed fraction of
the cluster gas, with the same fraction as is observed in the Coma cluster, $M_{B} = M_{gas}(1+0.20 h^{3/2})$.  For a realistic
equation of state, the gas in the cluster will be more extended than
the dark matter and the baryon fraction at
$r_{500}$ will be a modest underestimate of the true baryon fraction \citep{evrard1997}, $f_g(r_{500}) = 0.85\times f_b({\rm universal})$.  These
assumptions lead to
\begin{equation} f_B=(f_g(1 + 0.2h^{3/2})/ 0.85). \end{equation}
Using this estimate of the baryon fraction, and $h=0.7$ in a flat
cosmology, in
Figure~\ref{fig:fitomega} we show our best estimate of \omegam\ as a
function of cosmology.  Thus we find our best estimate of \omegam\ is $\sim$0.25.

\subsection{Future Work}
There are several improvements to this work which will be made in the
near future.  More clusters will be added to the sample as SZ
observations continue.  And the potential also exists for improving
the centimeter-wave SZ interferometer system
dramatically by taking advantage of the 10 GHz output of the SZ receivers; currently a maximum of 2 GHz are correlated at OVRO and effective
bandwidth of 0.5 GHz are correlated at BIMA.  

One of the main sources of uncertainty in these
measurements originates in the emission-weighted gas
temperature measurements; as $f_g\propto T_e^{-2}$, the 10-20\% uncertainties in
$T_e$ roughly double to 20-40\% uncertainties in the gas mass
fraction.  A large number of these clusters are scheduled to be
observed in the Chandra X-Ray Observatory GTO and GO phases, which
should improve the situation considerably.  

Numerical simulations will also help identify other sources of
systematic error incurred in the observational and analysis program.  
An analysis is in preparation of hydrodynamic simulations of a sample of clusters to
quantify any biases we may introduce to the gas mass fraction
measurements with the interferometric 
method and through the assumptions we make in the fitting and analysis
of the clusters.

        Many thanks are due to the staff at the BIMA and OVRO
observatories for their contributions to this project, especially Rick
Forster, John Lugten, Steve Padin, Dick Plambeck, Steve Scott, and
Dave Woody.  Many thanks to Cheryl Alexander for her work on the
system hardware.  
This work is supported by NASA LTSA grant NAG5-7986. 
LG, EDR, and SKP gratefully thank the NASA GSRP program for its
support.  JJM is supported by Chandra Fellowship grant PF8-1003,
awarded through the Chandra Science Center.  The Chandra Science
Center is operated by the Smithsonian Astrophysical Observatory for
NASA under contract NAS8-39073.  Radio astronomy with the OVRO and BIMA
millimeter arrays is supported by NSF grants AST 96-13717 and AST
96-19938, respectively.
The funds for the additional hardware for the SZ experiment were from
a NASA CDDF grant, a NSF-YI Award, and the David and Lucile Packard Foundation.
\newpage
\bibliographystyle{apj}
\bibliography{clusters,apjourn}

\clearpage
\begin{deluxetable}{lccccccc}
\tablenum{1}
\tablecolumns{8}
\tablewidth{0pc}
\tablecaption{The Cluster Sample\label{table:sample}}
\setlength{\tabcolsep}{1.5mm}
\renewcommand{\arraystretch}{1.60}
\tablehead{
\colhead{Cluster} & \colhead{z} & \colhead{reference} &
\colhead{$T_e$} & \colhead{reference} & \colhead{$L_x$} & \colhead{band} & \colhead{reference}\\
\colhead{} & \colhead{} & \colhead{} & \colhead{(keV)} & \colhead{} & \colhead{($10^{45}$ erg/s)}
& \colhead{(keV)} & \colhead{}
}
\startdata
Abell 2218 & 0.171 & LB & 7.05$^{+0.36}_{-0.35}$ & AF & 1.08 & 2-10 & AF \\
Abell 1914 & 0.1712 & BA & 10.7$^{+1.5}_{-1.5}$ & EB & 1.8 & 0.3-3.5 & EB \\
Abell 665 & 0.1818 & SR & 9.03$^{+0.58}_{-0.52}$ & AF & 1.78 & 2-10 & AF \\
Abell 1689 & 0.1832 & SR & 10.0 $^{+1.2}_{-0.80}$ & AF & 3.24 & 2-10 & AF \\
Abell 2261 & 0.224 & C95 & 10.09 $^{+5.9}_{-2.2}$ & AF & 2.39 & 2-10 & AF \\
Abell 1835 & 0.2528 & A92 & 9.8 $^{+2.3}_{-1.3}$ & AF & 4.54 & 2-10 & AF \\
Abell 697 & 0.282 & C95 & 9.80 $^{+0.70}_{-0.70}$ & AF & 1.574 & 0.1-2.4 & E98 \\
Abell 611 & 0.288 & C95 & 6.60 $^{+0.60}_{-0.60}$ & AF & 1.04 & 0.1-2.4 & E98 \\
Abell 1995 & 0.3219 & PA & 8.59$^{+0.86}_{-0.67}$ & MS & 0.87 & 0.3-3.5 & MS \\
ZwCl 1953 & 0.32 & BA & 13.2$^{+2.0}_{-2.0}$ & E98 & 2.86 & 0.3-3.5 & E98 \\
MS1358.4+6245 & 0.328 & GI & 7.48 $^{+0.83}_{-0.70}$ & AF & 1.08 & 2-10& AF \\
RXJ 1532.9+3021 & 0.345 & EB & 12.20$^{+2.00}_{-2.00}$ & E98 & 2.374 & 0.3-3.5 & E98 \\
Abell 370 & 0.374 & M88 & 6.60 $^{+1.10}_{-0.90}$ & OT & 1.3 & 2-10 & AE \\
CL0016+16 & 0.5479 & GI & 7.55$^{+0.72}_{-0.58}$ & HB & 1.46 & 0.3-3.5 & GI \\
MS0451.6-0305 & 0.55 & GI & 10.17$^{+1.55}_{-1.26}$ & MS & 0.7 & 0.3-3.5 & GI \\
MS2053.7-0449 & 0.583 & GI & 6.60 $^{+2.00}_{-2.00}$ & AEest & 0.58 & 0.3-3.5 & GI \\
MS1137.5+6625 & 0.78 & GI & 5.70 $^{+2.10}_{-1.10}$ & D99 & 1.9 & 0.3-3.5 & GI \\
MS1054.5-0321 & 0.826 & LG & 12.30$^{+3.10}_{-2.20}$ & D98 & 9.3 & 0.3-3.5 & LG \\
\enddata
\tablerefs{A92 Allen (1992); AF Allen \& Fabian (1998);
AE Arnaud \& Evrard (1998); AEest estimated from $L_x$-T relation of AE;
BA Bade, N. \etal\ (1998);  C95 Crawford \etal\ (1995);
D98 Donahue \etal\ (1998); D99 Donahue \etal\ (1999);
EB Ebeling \etal\ (1996), E98 Ebeling (1998);\
GI Gioia \etal\ (1990); HB Hughes \& Birkinshaw (1998);
 LB LeBorgne \etal\ (1992);
LG Luppino \& Gioia (1995); M88 Mellier \etal\ (1988);
MS Mushotzky \& Scharf (1997); OT Ota \etal\ (1998); PA Patel \etal\
(2000); SR Struble \& Roo
d (1991)}
\end{deluxetable}

\clearpage
\begin{deluxetable}{lcccccc}
\setlength{\tabcolsep}{1.5mm}
\renewcommand{\arraystretch}{1.60}
\tablewidth{0pt}
\tablecolumns{5}
\tablenum{2}
\tablecaption{SZ-derived Gas Masses and Mass Fractions, using
$\Omega_M=0.3$, $\Omega_\Lambda=0.7$\label{table:fgcirc}}
\tablehead{
\colhead{Cluster} &   \colhead{gas mass within 65$''$} & \colhead{$f_g h$\ (within\ 65$'') $}&  \colhead{$\underline{f_g(r_{500})}$}  & \colhead{$ f_g h$\ (within\ $r_{500})$}\\
\colhead{}& ($h^2/10^{12} M_\circ$) &\colhead{} &$f_g(65'')$&\colhead{}
}
\startdata
Abell 2218  & 4.91 $^{+1.39}_{-1.76}$& 0.179$^{+0.037}_{-0.046}$ & 1.40 & 0.250$^{+0.051}_{-0.065}$ \\
Abell 1914  & 3.51 $^{+1.04}_{-1.05}$& 0.037$^{+0.019}_{-0.019}$ & 1.45 & 0.053$^{+0.027}_{-0.027}$ \\
Abell 665   & 1.97 $^{+0.67}_{-0.54}$& 0.042$^{+0.022}_{-0.022}$ & 1.42 & 0.060$^{+0.031}_{-0.031}$ \\
Abell 1689  & 4.60 $^{+0.82}_{-1.14}$& 0.068$^{+0.020}_{-0.023}$ & 1.43 & 0.098$^{+0.029}_{-0.033}$ \\
Abell 2261  & 3.12 $^{+2.74}_{-3.76}$& 0.027$^{+0.070}_{-0.016}$ & 1.39 & 0.037$^{+0.097}_{-0.022}$ \\
Abell 1835  & 4.82 $^{+1.33}_{-2.31}$& 0.085$^{+0.026}_{-0.042}$ & 1.36 & 0.116$^{+0.035}_{-0.057}$ \\
Abell 697   & 3.66 $^{+2.01}_{-0.71}$& 0.021$^{+0.043}_{-0.006}$ & 1.34 & 0.029$^{+0.057}_{-0.009}$ \\
Abell 611   & 5.05 $^{+4.11}_{-1.37}$& 0.048$^{+0.140}_{-0.024}$ & 1.29 & 0.062$^{+0.180}_{-0.030}$ \\
ZwCl 1953  & 3.23 $^{+1.81}_{-1.43}$& 0.054$^{+0.019}_{-0.027}$ & 1.34 & 0.073$^{+0.026}_{-0.036}$ \\
Abell 1995  & 7.44 $^{+1.69}_{-1.92}$& 0.079$^{+0.030}_{-0.031}$ & 1.30 & 0.102$^{+0.039}_{-0.041}$ \\
MS1358.4+6245 & 6.00 $^{+1.59}_{-2.17}$& 0.097$^{+0.067}_{-0.049}$ & 1.28 & 0.124$^{+0.086}_{-0.062}$ \\
RXJ 1532.9+3021  & 4.80 $^{+1.85}_{-1.71}$& 0.038$^{+0.030}_{-0.016}$ & 1.32 & 0.050$^{+0.040}_{-0.021}$ \\
Abell 370   & 8.57 $^{+2.60}_{-2.90}$& 0.087$^{+0.045}_{-0.048}$ & 1.24 & 0.108$^{+0.055}_{-0.059}$ \\
CL0016+16 & 18.85$^{+4.64}_{-4.11}$& 0.139$^{+0.086}_{-0.039}$ & 1.19 & 0.165$^{+0.103}_{-0.046}$ \\
MS0451.6-0305 & 21.13$^{+5.62}_{-7.98}$& 0.128$^{+0.050}_{-0.055}$ & 1.22 & 0.155$^{+0.061}_{-0.068}$ \\
MS2053.7-0449 & 6.44 $^{+7.21}_{-4.19}$& 0.044$^{+0.136}_{-0.034}$ & 1.16 & 0.052$^{+0.158}_{-0.040}$ \\
MS1137.5+6625 & 15.75$^{+6.33}_{-11.74}$& 0.062$^{+0.037}_{-0.048}$ & 1.10 & 0.068$^{+0.041}_{-0.053}$ \\
MS1054.5-0321 & 14.18$^{+5.27}_{-7.28}$& 0.045$^{+0.024}_{-0.024}$ & 1.17 & 0.053$^{+0.028}_{-0.028}$ \\
\enddata
\end{deluxetable}

\clearpage

\begin{deluxetable}{cccccccc}
\tablewidth{0pt}
\tablenum{3}
\setlength{\tabcolsep}{1.5mm}
\renewcommand{\arraystretch}{1.60}
\tablecaption{Mean Gas Mass Fractions\label{table:results}}
\tablecomments{The mean gas mass fractions for the noted samples are
presented.  The gas mass fractions depend on $D_A^{-1}$ and so the
results are presented for three sets of cosmological parameters.
($H_\circ = 100 h$ km s$^{-1}$ Mpc$^{-1}$).}
\tablehead{
\colhead{}&\colhead{}&\multicolumn{2}{c}{$\Omega_M$=0.3,$\Omega_\Lambda$=0.7}&\multicolumn{2}{c}{$\Omega_M$=0.3,$\Omega_\Lambda$=0.0}&\multicolumn{2}{c}{$\Omega_M$=1.0,$\Omega_\Lambda$=0.0}\\
\cline{3-4} \cline{5-6} \cline{7-8} \\
\colhead{sample}& \colhead{number of
clusters}&\colhead{$\overline{f_g h_{100}}$}&\colhead{$\chi^2_{red}$}&\colhead{$\overline{f_g
h_{100}}$}&\colhead{$\chi^2_{red}$}&\colhead{$\overline{f_g h_{100}}$}&\colhead{$\chi^2_{red}$}
}

\startdata
full sample&18&$0.081^{+0.009}_{-0.011}$&1.021&$0.074^{+0.008}_{-0.009}$&1.027&$0.068^{+0.009}_{-0.008}$&1.056 \\
EMSS subsample&6&$0.089^{+0.018}_{-0.019}$&1.208&$0.077^{+0.017}_{-0.016}$&1.258&$0.067^{+0.015}_{-0.014}$&1.352\\
\enddata
\end{deluxetable}

\clearpage

\clearpage

\clearpage
\clearpage
\begin{figure}
\centerline{\hbox{
\psfig{figure=fig1b.epsi,angle=270}}}
\end{figure}

\clearpage

\begin{figure}
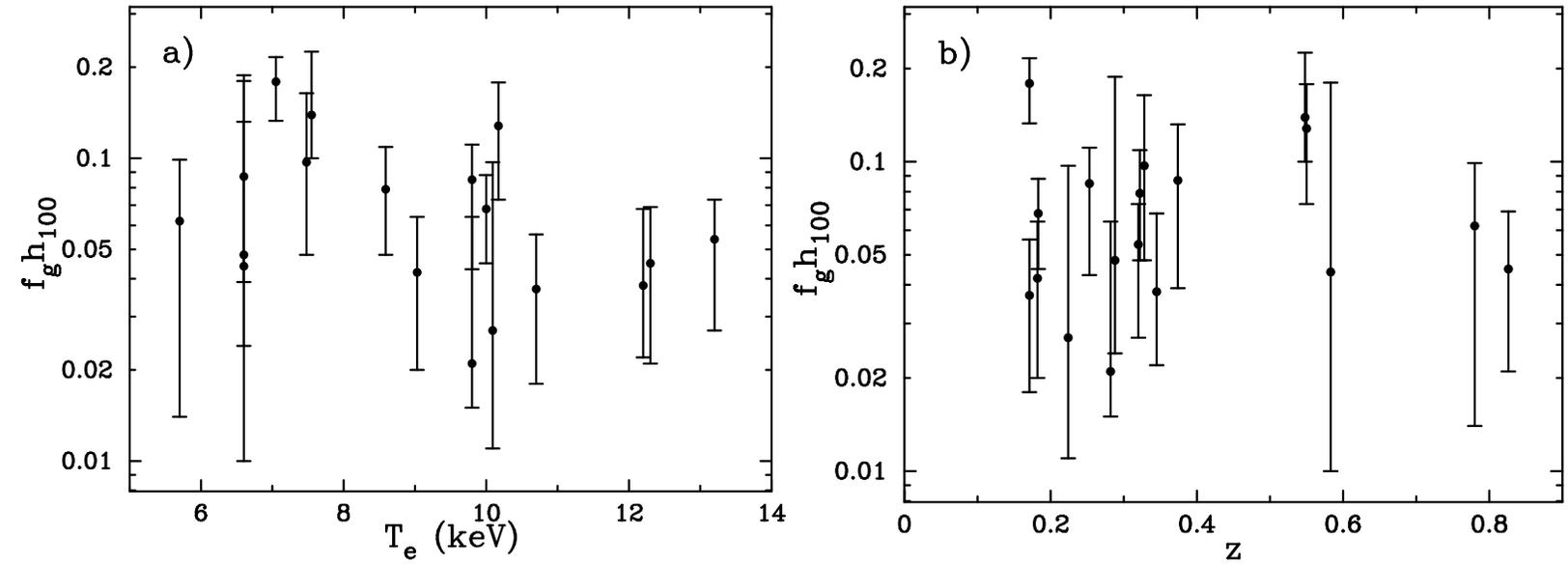

\figurenum{3}
\epsscale{1.0}
\centerline{\hbox{
\psfig{figure=fig3a.epsi,height=3.0in,angle=270}
\psfig{figure=fig3b.epsi,height=3.0in,angle=270}
}}
\caption{\label{fig:circ_fgvz,T} Gas mass fractions within $r_{500}$
for the entire sample, assuming the cosmology (\omegam=0.3, \omegal=0.7). a) Gas mass fraction vs. ICM emission-weighted temperature. b) Gas mass fraction versus redshift.}
\label{fig3}
\end{figure}

\clearpage

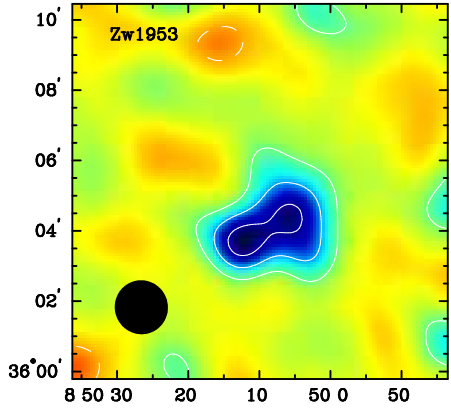
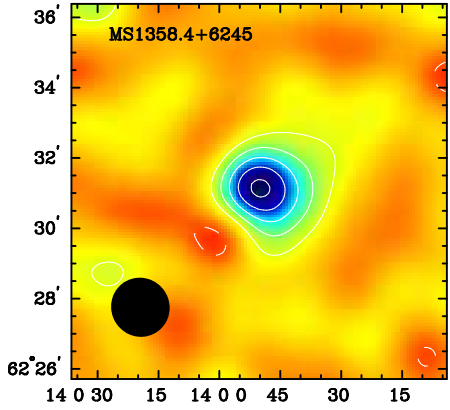
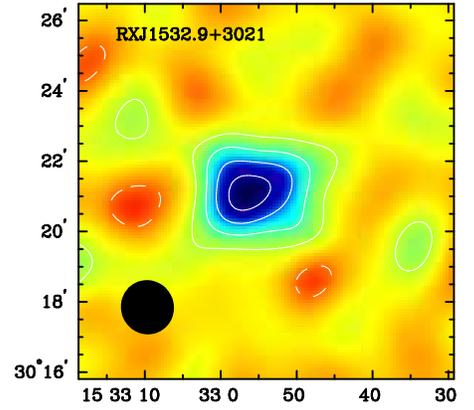
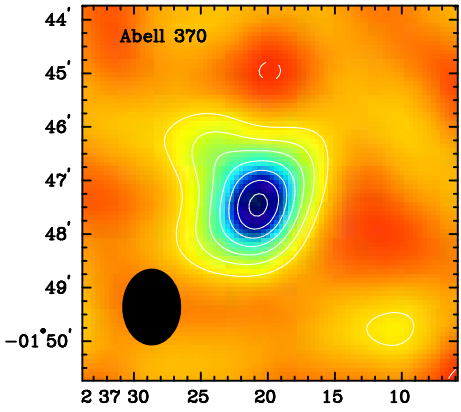
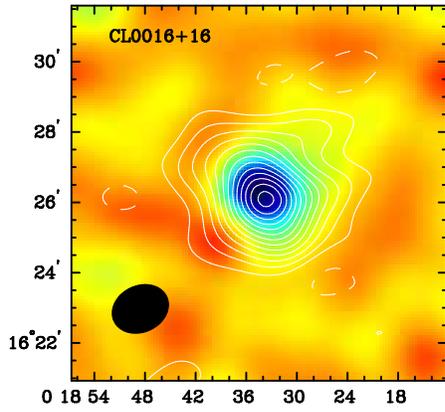
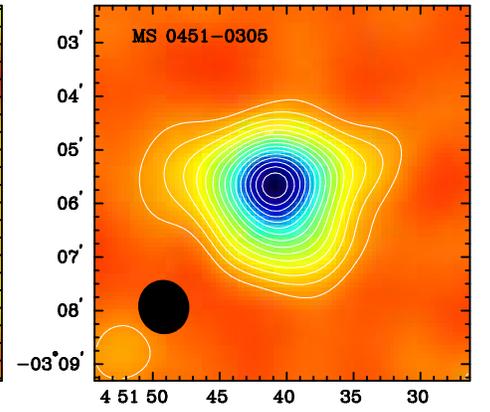
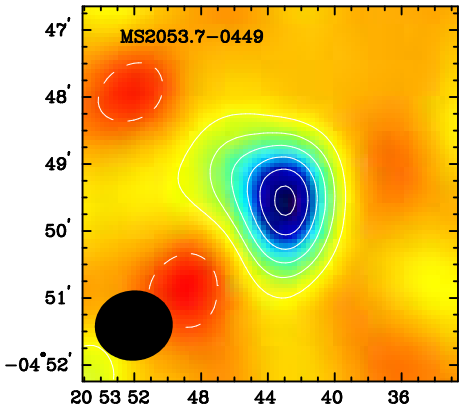
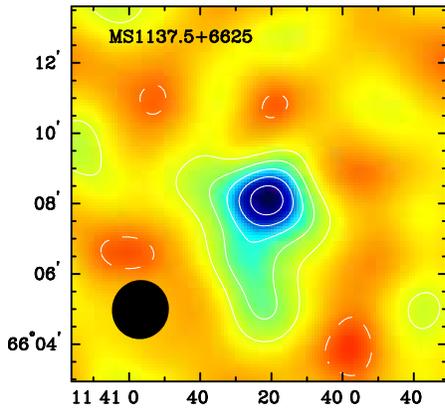
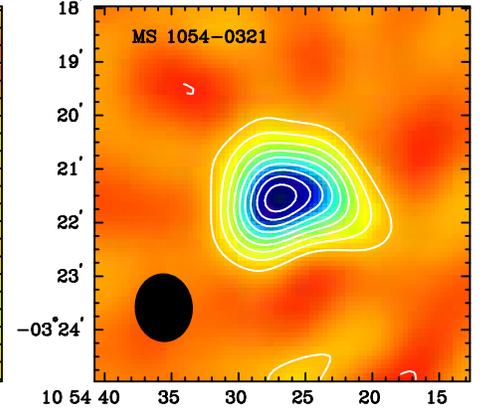
\end{document}